\documentclass[aps,prl,twocolumn,psfig,showpacs,superscriptaddress]{revtex4-1}
\usepackage{amsfonts}
\usepackage{dsfont}
\usepackage{amsmath}
\usepackage{graphicx}
\usepackage{bm}
\usepackage{amssymb}
\usepackage{times}
\usepackage{dcolumn}
\usepackage{cases}
\usepackage{txfonts}
\usepackage{color}
\RequirePackage[hyperindex,colorlinks,bookmarksnumbered,plainpages=true]{hyperref}
\hypersetup{colorlinks,linkcolor=blue,urlcolor=blue,citecolor=blue}
\usepackage{hyperref}
\usepackage{ulem}%only for the command \sout = scrap

\begin{document}
\title{Identifying Criticality in Higher Dimensions by Time Matrix Product State}

\author{Cheng Peng}
\affiliation{Theoretical Condensed Matter Physics and Computational Materials Physics Laboratory, School of Physical Sciences, University of Chinese Academy of Sciences, Beijing 100049, China}
\author{Shi-Ju Ran}
\email[Corresponding author. ]{Email: shi-ju.ran@icfo.eu}
\affiliation{ICFO-Institut de
	Ciencies Fotoniques, The Barcelona Institute of Science and
	Technology, 08860 Castelldefels (Barcelona), Spain}
\author{Maciej Lewenstein}
\affiliation{ICFO-Institut de
	Ciencies Fotoniques, The Barcelona Institute of Science and
	Technology, 08860 Castelldefels (Barcelona), Spain}
\affiliation{ICREA, Lluis Companys 23, 08010 Barcelona, Spain}
\author{Gang Su}
\email[Corresponding author. ]{Email: gsu@ucas.ac.cn}
\affiliation{Theoretical Condensed Matter Physics and Computational Materials Physics Laboratory, School of Physical Sciences, University of Chinese Academy of Sciences, Beijing 100049, China}
\affiliation{Kavli Institute for Theoretical Sciences, University of Chinese Academy of Sciences, Beijing 100190, China}

\begin{abstract}
  Characterizing criticality in quantum many-body systems of dimension $\ge 2$ is one of the most important challenges of the contemporary physics. In principle, there is no generally valid theoretical method that could solve this problem. In this work, we propose an efficient approach to identify the criticality of quantum systems in higher dimensions. Departing from the analysis of the numerical renormalization group flows, we build a general equivalence between the higher-dimensional ground state and a one-dimensional (1D) quantum state defined in the imaginary time direction in terms of the so-called time matrix product state (tMPS). We show that the criticality of the targeted model can be faithfully identified by the tMPS, using the mature scaling schemes of correlation length and entanglement entropy in 1D quantum theories. We benchmark our proposal with the results obtained for the Heisenberg anti-ferromagnet on honeycomb lattice. We demonstrate critical scaling relation of the tMPS for the gapless case,  and a trivial scaling for the gapped case with spatial anisotropy. The critical scaling behaviors are insensitive to the system size, suggesting the criticality can be identified in small systems. Our tMPS scheme for critical scaling shows clearly that the spin-1/2 kagom\'e Heisenberg antiferromagnet has a gapless ground state. More generally, the present study indicates that the 1D conformal field theories in imaginary time provide a very useful tool to characterize the criticality of higher dimensional quantum systems.
\end{abstract}

\maketitle

\textit{Introduction---}. The characterization of phases of matter and phase transition belongs to the most fundamental problems in physics. This problem is particularly challenging  for the phases, for which the Landau-Ginzburg paradigm fails.  Such phases  possess no local orders or broken spontaneous symmetries; prominent examples include  fractional quantum Hall states \cite{STG99HallRev}, or quantum spin liquids (QSL's) \cite{M00QSLrev,B10QSLRev,SB17QSLRev,ZKN17QSLrev}. One powerful tool that describes such elusive many-body systems is the conformal field theory (CFT) \cite{FMS99CFTBook,HLW94CFTRG}. Combining CFT with methods in quantum information science and strongly correlated systems, the scaling schemes for determining the conformal features of the critical one-dimensional (1D) quantum systems were proposed \cite{VLRK03CritEnt,TOIL08EntScaling,PMTM09EntScaling,SHMTV15ScalingcMPS,WGK17EntScaling}. A general idea is to connect the entropy of the CFT with the quantum entanglement \cite{BBPS95EntCFT0}, as it is done  in spin systems. In general, the behaviors of systems in 1D  (and for small systems even in 2D) are studied by the powerful numerical state ansatz, called matrix product state (MPS) \cite{W92DMRG,W93DMRG,M07DMRGsymme,S11DMRGRev}.

In 1D, the criticality has been well understood, thanks to the efficient algorithms such as density matrix renormalization group (DMRG) \cite{W92DMRG,W93DMRG}, or time evolving block decimation (TEBD)\cite{V04TEBD,V07iTEBD}. The criticality of the ground states of 1D quantum models can be efficiently determined, and the central charge can be accurately obtained by the scaling of the correlation length and entanglement entropy, when the ground state is critical \cite{TOIL08EntScaling}.

In higher dimensions, however, due to the lack of CFT and due to the complexities of numerical approaches, many issues about the criticality are still unsettled. One paradigmatic model is the spin-1/2 Heisenberg antiferromagnet (HAFM) on kagom\'e lattice: It is still under a very hot debate, whether its ground state is gapless \cite{INBSN08KagomeExp,HMSNB+07kagomeexp,OMBDTV+08KagomeExp,WLSRM+10kagomeexp,HHCNR+12fractionalized,HCBFZH+14KagomeGplessExp,RMAF07KagomeGapless,RHLW07kagomeQSL,HRLW08aQSL,SL09LowEnegExcKagome,IBSP13kagomeQSL,IPB14kagomeQSL,LXCLXH+17kagomeQSL,CRLPHS17arxiv,HZOP17kagomeDMRG}, or gapped \cite{IFHL11KagomeGappedExp,WEBLSL+98KagomeGapped,JWS08DMRGKagome,HT10KagomeGapped,DMS12DMRGKagome,YHW12DMRGkagome,JWB12TopoEntKagome,NSH12kagomemag,CDHLR13KagomePlateau}. The existing methods that are frequently used to identify the criticality require very high precision of numerical approach to capture, e.g., the long-range correlations or low-temperature thermodynamics (such as specific heat).  These  quantities and behaviors are very challenging to access with the current numerical techniques. The scaling schemes in higher dimensions are not well established. Moreover, it is also unclear, how to capture a critical state in an efficient state ansatz such as projected entangled pair state (PEPS) \cite{RPLLS17Scaling2D}. One alternative method is the multiscale entanglement renormalization ansatz  (MERA) \cite{V07EntRenor} that introduces an extra dimension of renormalization to capture the criticality \cite{PEV09MERAcrit,MRGF09critMERA,EV13MERAbook}; despite many successful applications, MERA method is still under development.

Recently, the time matrix product state (tMPS) \cite{TLR16tMPSarxiv} was proposed in infinite 1D quantum lattice models with translational invariance. Using the Trotter-Suzuki decomposition \cite{SI87Trotter,IS88Trotter}, the 1D quantum model is mapped onto a two-dimensional (2D) tensor network (TN), where the tMPS is defined as a 1D state along the imaginary time. The physical properties of the ground state can be accurately obtained from the tMPS. In particular in the gapless case, the tMPS exhibits the critical features such as a logarithmic correction to area law of the entanglement entropy, an algebraic scaling of the correlation length, and a central charge that characterizes the criticality.

In this work, we generalize the tMPS to ($\mathcal{D}>1$)--dimensional quantum systems with the help of numerical renormalization group (NRG) flows, and show that the criticality can be accurately identified by the tMPS. Since the tMPS is always 1D regardless of the dimensions of the model, the idea is to use the scaling theories for 1D critical quantum states to characterize the criticality of higher-dimensional systems through the tMPS. With the effective basis defined by the NRG flows, we build the equivalence between the tMPS and the ground state for the correlation length and entanglement, providing a mathematical framework to understand the connections between these two states.

%To tackle the difficulties brought by the high dimensionality, we choose a finite-size system with the periodic boundary condition.

We test our scheme on the isotropic HAFM on honeycomb lattice whose ground state is believed to be gapless \cite{RRY892DHAF}. The tMPS exhibits the critical scaling behaviors of the 1D quantum states, giving a logarithmic relation between the entanglement entropy and correlation length \cite{VLRK03CritEnt,TOIL08EntScaling}. Our results show that the criticality that appears in the thermodynamic limit can be faithfully captured by the tMPS on the systems of moderate small sizes. This is in contrast to the existing methods for higher-dimensional systems, which suffer from strong finite-size effects. We apply then our scheme to the spin-1/2 HAFM on kagom\'e lattice, showing that the ground state is a gapless QSL.

%Meanwhile, the scaling of system size are extremely expensive in higher dimensions.

%Criticality of PEPS is unclear (like the argument in our scaling PRB). No efficient approach to determine the 2D criticality.

% It's hard to observe the criticality on 2D lattice, e.g., whether the ground state of Heisenberg antiferromagnetic model on Kagome lattice is a gapped $\mathds{Z}_2$ spin-liquid or a gapless critical state still remaining a hot debatable issue. The reason of the difficulty comes from that the finite size effect on lattice is much stronger than that of 1D system especially for frustrated and critical system. In case of that a theory of grasping criticality of 2D systems is in urgent need. As we have already known, a serious of scaling theory in 1D which projects the $1$-D quantum model to $(1+1)$-D equivalent system constructed using continues MPS (cMPS) along imaginary time direction to investigate the criticality is widely accepted. The characters captured by cMPS are robust and free from finite size effect. In this paper, we would introduce tMPS in 2d to investigate the criticality, and our method shows the robust 2D conformal field property under small system size.

\textit{Numerical renormalization group flow.---} To present our theory, we consider a 1D $N$-site system with the periodic boundary condition. The Hamiltonian is $\hat{H}=\sum_{n}\hat{H}^{[n,n+r]}$, where $\hat{H}^{[n,n+r]}$ gives the two-body interactions with $r$ the coupling distance. Note that most of the arguments below can be directly applied to any models of finite sizes in ($\mathcal{D} \geq1$) dimensions without substantial changes.

\begin{figure}
	\begin{minipage}{\linewidth}
		\includegraphics[angle=0,width=\textwidth]{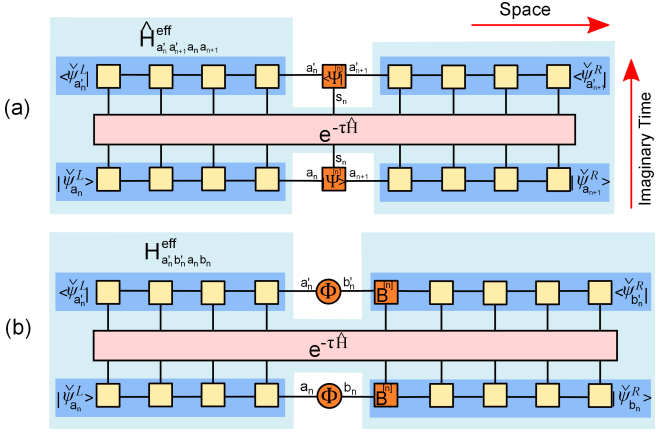}
	\end{minipage}
	\caption{\label{Heff}
    (Color online) (a) Graphical representation of $\sum_{a_n'a_{n+1}' a_na_{n+1}} \langle \Psi^{[n]}_{a_n'a_{n+1}'}| \hat{H}^{eff}_{a_n'a_{n+1}' a_na_{n+1}} |\Psi^{[n]}_{a_na_{n+1}} \rangle$. We use the pink rectangle to represent $e^{-\tau \hat{H}}$. The yellow squares are the local tensors $A$ and $B$ in Eqs. (\ref{eq-RGflow1}) and (\ref{eq-RGflow2}). The blue shadows are the NRG flows $|\check{\psi}^L_{a_{n}}\rangle$ and $|\check{\psi}^R_{a_{n+1}}\rangle$. The orange square represents $|\Psi^{[n]}_{a_na_{n+1}} \rangle$, which is the ``ground state'' of the effective Hamiltonian $\hat{H}^{eff}_{a_n'a_{n+1}' a_na_{n+1}}$ (depicted by light blue part) in Eq. (\ref{eq-Heff}). The space and the imaginary time directions are shown by red arrows. (b) The QR decomposition of $|\Psi^{[n]}_{a_na_{n+1}} \rangle$ and the redefinition of $H^{eff}_{a_n'b_{n}', a_nb_{n}}$ in Eq. (\ref{eq-Heffnew}).
    }
\end{figure}

When $N$ is large, exact solution the ground state is essentially impossible, since the Hilbert space increases exponentially with the system size. Luckily, for many models under interests, the low-energy states satisfy the so-called area law of entanglement entropy \cite{ECP10AreaLawRev}, and the valid Hilbert space only lies in a very small corner of the full space of many body states. In this case, the many-body system can be faithfully described by an effective model defined only in the relevant Hilbert space.

Using the NRG based schemes, e.g., DMRG \cite{W92DMRG,W93DMRG}, such a finite model is described by an effective Hamiltonian under the NRG flows towards a certain (say, the $n$-th) site (Fig.~\ref{Heff}). The effective Hamiltonian can be formally written as
\begin{equation}
\hat{H}^{eff}_{a_n'a_{n+1}' a_na_{n+1}}= \langle \check{\psi}^L_{a_n'}| \otimes \langle \check{\psi}^R_{a_{n+1}'}| e^{-\tau \hat{H}} |\check{\psi}^L_{a_n}\rangle \otimes |\check{\psi}^R_{a_{n+1}}\rangle.
\label{eq-Heff}
\end{equation}
$|\check{\psi}^L_{a_n}\rangle$ (and $|\check{\psi}^R_{a_{n+1}}\rangle$) represents the NRG flow that renormalizes the Hilbert space of the sites on the left (and right) side of the site $n$, to an effective space denoted by $a_n$ (and $a_{n+1}$). Utilizing the MPS representation, $|\check{\psi}^L_j\rangle$ can be written as
\begin{eqnarray}
|\check{\psi}^L_{a_n}\rangle &=& \sum_{s_1\cdots s_{n-1}} \sum_{a_1\cdots a_{n-1}} A^{[1]}_{s_{1},a_{1}a_{2}} \cdots A^{[n-1]}_{s_{n-1},a_{n-1}a_{n}} |s_1 \cdots s_{n-1} \rangle,\label{eq-RGflow1}\\
|\check{\psi}^R_{a_{n+1}}\rangle &=& \sum_{s_{n+1}\cdots s_{N}} \sum_{a_{n+2}\cdots a_{N+1}} B^{[n+1]}_{s_{n+1},a_{n+1}a_{n+2}} \cdots B^{[N]}_{s_N,a_{N}a_{N+1}} |s_{n+1} \cdots s_{N} \rangle, \nonumber.\label{eq-RGflow2}\\
\end{eqnarray}
with $|s_n\rangle$ representing the local basis of the $n$-th site.
$A^{[n]}$ and $B^{[m]}$ are ($d\times \chi \times \chi$) tensors defined on the corresponding sites. Here $d$ is the dimension of a physical site, while  $\chi$ is the dimension cut-off of the virtual indices $\{a_n\}$, which characterizes the maximum of the entanglement that the NGR flows can carry. We take the dimensions of $a_{1}$ and $a_{N+1}$ equal to one, since the NRG flows  $|\check{\psi}^L_{a_n}\rangle$ and $|\check{\psi}^R_{a_{n+1}}\rangle$ start from the 1st and the $N$-th sites, respectively. To define the NRG flows, these tensors satisfy the following orthogonal conditions
\begin{eqnarray}
\sum_{s_na_n} A^{[n]\ast}_{s_{n},a_{n}a_{n+1}'} A^{[n]}_{s_{n},a_{n}a_{n+1}} = I_{a_{n+1}'a_{n+1}}, \label{eq-Ort1} \\
\sum_{s_ma_{m+1}} B^{[m]\ast}_{s_{m},a_{m}'a_{m+1}} B^{[m]}_{s_{m},a_{m}a_{m+1}} = I_{a_{m}'a_{m}}. \label{eq-Ort2}
\end{eqnarray}

Here, we choose the exponential form $e^{-\tau \hat{H}}$ to define the effective Hamiltonian so that we can apply the Trotter-Suzuki scheme, where $\tau$ is a positive small number called Trotter-Suzuki step. In our calculations $\tau$ is as small as $10^{-6}$, and $e^{-\tau \hat{H}}$ is approximated by $I-\tau \hat{H}$ for a higher efficiency with DMRG.

The tensors in Eqs. (\ref{eq-RGflow1}) and (\ref{eq-RGflow2}) can be optimally determined using the DMRG algorithm. One may refer to Refs. [\onlinecite{W92DMRG,W93DMRG}] for more details. The ground state $|\psi\rangle$ is then given by an MPS that reads
\begin{equation}
|\psi\rangle = \sum_{a_na_{n+1}} |\check{\psi}^L_{a_n}\rangle  |\Psi^{[n]}_{a_na_{n+1}} \rangle |\check{\psi}^R_{a_{n+1}}\rangle,
\label{eq-GS}
\end{equation}
where $|\Psi^{[n]}_{a_na_{n+1}} \rangle = \sum_{s_n} \Psi^{[n]}_{s_n,a_na_{n+1}} |s_n \rangle$ is the ``ground state'' of the effective Hamiltonian in Eq. (\ref{eq-Heff}). We shall stress that the NRG flow is considered here to be concept independent of the algorithms, in particular, it could be obtained by methods other than DMRG. Our scheme is applicable to models of a finite size, but of an arbitrary dimension, as long as the NRG flows accurately capture the valid Hilbert space.

% ($d\chi^2 \times d\chi^2$) matrix $H^{eff}_{s_n'a_n'a_{n+1}', s_na_na_{n+1}} = \langle s_n'| \hat{H}^{eff}_{a_n'a_{n+1}' a_na_{n+1}} |s_n \rangle$.

\textit{Time matrix product state in finite density matrix renormalization group.---}
It has been shown \cite{R16AOP,TLR16tMPSarxiv} that in an infinite 1D system, when the DMRG defines the NRG flows in the real space, tMPS appears in the imaginary time direction, and corresponds to the infinite time-evolving block decimation procedure \cite{V07iTEBD}. In the following, we show how to define the tMPS in the DMRG scheme of a finite system in any dimensions.

To this aim, we modify the effective Hamiltonian, so that the NRG flows meet at the virtual index $a_n$. We introduce the tensor $B^{[n]}$ on the $n$-th site by the QR decomposition of $|\Psi^{[n]}_{a_na_{n+1}} \rangle$ as
\begin{equation}
 |\Psi^{[n]}_{a_na_{n+1}} \rangle = \sum_{b_n} (\Phi_{a_nb_n} \sum_{s_n} B^{[n]}_{s_n, b_na_{n+1}} |s_n\rangle).
 \label{eq-Bn}
\end{equation}
One can see that $B^{[n]}$ satisfies the orthogonal condition in Eq. (\ref{eq-Ort2}). Then, the NRG flow $|\check{\psi}^R_{a_{n+1}}\rangle$ is modified by adding the $n$-th site as
\begin{equation}
|\check{\psi}^R_{b_{n}}\rangle = \sum_{s_{n}\cdots s_{N}} \sum_{a_{n+1}\cdots a_{N+1}} B^{[n]}_{s_{n},b_{n}a_{n+1}} \cdots B^{[N]}_{s_N,a_{N}a_{N+1}} |s_{n+1} \cdots s_{N} \rangle.
\label{eq-flowR1}
\end{equation}

By constructing in a similar way as Eq. (\ref{eq-Heff}), the new effective Hamiltonian satisfies
\begin{equation}
H^{eff}_{a_n'b_{n}', a_nb_{n}}= \langle \check{\psi}^L_{a_n'}| \otimes \langle \check{\psi}^R_{b_{n}'}| e^{-\tau \hat{H}} |\check{\psi}^L_{a_n}\rangle \otimes |\check{\psi}^R_{b_{n}}\rangle,
\label{eq-Heffnew}
\end{equation}
which contains no physical indices, and is simply a ($\chi^2 \times \chi^2$) matrix. In fact, the ``ground state'' of the new effective Hamiltonian in Eq. (\ref{eq-Heffnew}) is the $\Phi$ in Eq. (\ref{eq-Bn}), and the ground state MPS can be written as
\begin{equation}
|\psi\rangle = \sum_{a_nb_n} |\check{\psi}^L_{a_n}\rangle  \Phi_{a_nb_n} |\check{\psi}^R_{b_n}\rangle.
\label{eq-GSnew}
\end{equation}
Compared with Eq. (\ref{eq-GS}), it gives exactly the same state up to a guage transformation on the index $a_n$ (or $b_n$).

Let us decompose $H^{eff}$ symmetrically as
\begin{equation}
H^{eff}_{a_n'b_n', a_nb_n}= \sum_{c_m} V^{\ast}_{c_m,a_n'a_n} V_{c_m,b_n'b_n}.
\label{eq-VLR}
\end{equation}
Here we assume such a decomposition exists, especially when the system has reflection symmetry of the canonical center. With the periodic boundary condition of the system, any site can be taken as the canonical center. The tMPS in our case is formed by infinite number of the copies of $V$ along the imaginary time direction as
\begin{equation}
\sum_{\{a\}}  \cdots V_{c_m,a_ma_{m+1}}  V_{c_{m+1},a_{m+1}a_{m+2}} \cdots.
\label{eq-tMPS}
\end{equation}
In the tMPS, $\{c\}$ play the role of the ``physical'' indices, and $\{a\}$ the virtual indices. See more details in the Appendix or Ref. [\onlinecite{TLR16tMPSarxiv}].

\textit{Equivalence of correlation length and entanglement in the renormalized basis.---}
Below, we show the equivalence of correlation length and entanglement between the ground state and the tMPS. Since the tMPS is a 1D quantum state and can in principle be studied by 1D quantum theories such as CFT, this equivalence provides us a theoretical ground to identify the criticality of higher-dimensional quantum models. The equivalence is based on the fact that Eq. (\ref{eq-VLR}) is both the effective Hamiltonian (in the exponential form) and the transfer matrix of the tMPS (shown in the Appendix).

Since the tMPS is translational invariant, its correlation length $\xi_T$ is given as $\xi_T = \frac{1}{\ln e_0 - \ln e_1}$, where $e_0$ and $e_1$ are the two leading eigenvalues of the transfer matrix of the tMPS. Meanwhile, the gap of the effective Hamiltonian gives the dynamic correlation length of the ground state. Thus, the correlation length is exactly the dynamic correlation length of the ground state in the renormalized basis.

For an infinite translational-invariant MPS, the entanglement spectrum $\Lambda$ can be obtained by canonicalization \cite{OV08canonical}. The guage transformation is obtained by the dominant eigenstate $\Phi$ of the transfer matrix, which is again the effective Hamiltonian in Eq. (\ref{eq-Heffnew}). Since in DMRG, the effective Hamiltonian is Hermitian, the entanglement spectrum of the tMPS in this case is exactly the singular value spectrum of $\Phi$ obtained by the singular value decomposition $\Phi_{a_nb_n} = \sum_{\alpha} U_{a_n  \alpha} \Lambda_{\alpha} D^{\ast}_{b_n \alpha}$. On the other hand, one can easily see from Eq. (\ref{eq-GSnew}) that this singular value spectrum is the entanglement spectrum of the ground state due to the orthogonal conditions in Eqs. (\ref{eq-Ort1}) and (\ref{eq-Ort2}). Thus, the entanglement spectrum of the ground state in the renormalized basis is equivalent to that of the tMPS.

We shall stress that the equivalence is based on how accurate the NRG flows can capture the valid Hilbert space. Another requirement is the existence of the symmetrical decomposition of the effective Hamiltonian [Eq. (\ref{eq-VLR})]. Our simulations imply that even if these requirements are not strictly satisfied (e.g., with limited truncation errors or a slight breaking of the reflection symmetry), the correlation length and entanglement entropy of the tMPS still accurately give the properties of the ground state.

\begin{figure}
    \begin{minipage}{\linewidth}
        \includegraphics[width=\textwidth]{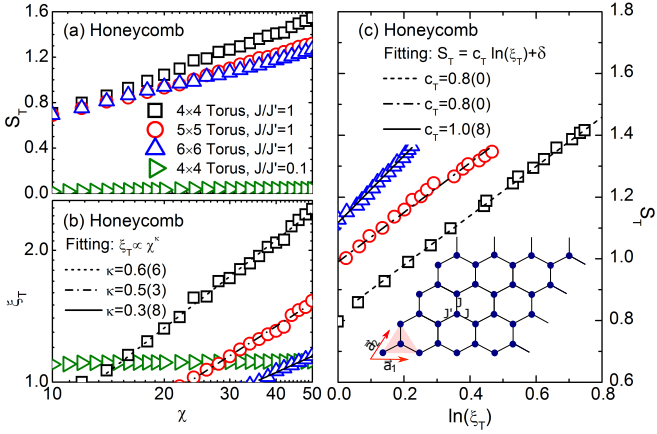}
    \end{minipage}
\caption{\label{HCScaling}
(Color online) The ground state scaling of (a) temporal entanglement entropy $S_T$ in the semi-log plot and (b) correlation length $\xi_T$ in the log-log plot of the spin-1/2 anisotropic HAFM on honeycomb lattice. The x-axis scales the dimension cut-off $\chi$ in DMRG. (c) The $S_T$-$ln(\xi_T)$ scaling. Inset: honeycomb lattice on $4 \times 4$ torus. The red shaded part is the unit cell with $\vec{a}_1$ and $\vec{a}_2$ the basis vectors. $J$ and $J'$ are the coupling constants. }
\end{figure}

\textit{Results and discussions.---}
We test our scheme on the spin-1/2 HAFM on an anisotropic honeycomb lattice. The Hamiltonian reads
\begin{equation}
\hat{H}=J\sum_{<ij>}\hat{S}_i\hat{S}_j+J'\sum_{<lm>}\hat{S}_l\hat{S}_m,
\label{eq6}
\end{equation}
where $J$ and $J'$ are the coupling constants of different directions as shown in the inset of Fig. \ref{HCScaling}(c). For $J/J'<J^c$ with $J^c=0.5(4)$, the ground state is a gapped dimerized state, while for $J^c<J/J'\leq 1$ it is a gapless semiclassical N\'eel state \cite{LGZS10honeycomb}.

We choose a finite lattice of the size ($L_1\vec{a}_1, L_2\vec{a}_2$) with the periodic boundary condition (hereinafter abbreviated as $L_1\times L_2$ torus), where $\vec{a}_1$ and $\vec{a}_2$ are the basis vectors shown in the inset of Fig. \ref{HCScaling}(c). We set $J/J'=0.1$ and $J/J'=1$ to show two different scaling behaviors (non-critical and critical) of the tMPS. Note that to remove as much as possible the effects brought by the deviation of the symmetrical requirement [Eq. (\ref{eq-VLR})], we calculate the quantities (entanglement entropy and correlation length) at every site and then take the average.

% We exhibit the finite size effect by computing $S_T$ and $\xi_T$ on $4 \times 4$ torus and $5 \times 5$ torus. It is perhaps no surprise to see the $\xi_T$ and $S_T$ fluctuate with the position. This is because the spatial translational invariance is broken in the finite size system. To do the scaling of $\xi_T$ and $S_T$, we computed $S_T$ and $\xi_T$ site by site and take the average of them.

As shown in Fig. \ref{HCScaling}, two significantly different behaviors of the entanglement entropy $S_T$ and the correlation length $\xi_T$ are found versus the dimension cut-off $\chi$ of the tMPS. Note $S_T$ is defined as $S_T = - \sum_{j} \Lambda_j^2 \ln \Lambda_j^2$. On the one hand, for $J/J'=0.1$, $S_T$ and $\xi_T$ saturate to a small value when $\chi$ increases. This coincides with the properties of a gapped 1D state, whose entanglement entropy and correlation length are finite.

On the other hand, for the isotropic point $J/J'=1$ where the ground state is critical, $S_T$ and $\xi_T$ increase along with increasing $\chi$, in contrast to the gapped case. By fitting, we find that $S_T$ gives a logarithmic relation and $\xi_T$ exhibits an algebraic law with $\chi$ as
\begin{eqnarray}
S_T = \alpha \ln \chi +const., \ \ \ \ \xi_T \propto \chi^{\kappa}.
\label{eq-scaling}
\end{eqnarray}
These are exactly the relations predicted by the ($1+1$)-dimensional CFT for critical 1D states \cite{VLRK03CritEnt,TOIL08EntScaling}. Our results suggest that the criticality of the ground state is faithfully given by the tMPS.

We also calculate the systems of different sizes at the isotropic point, where the scaling relations of $\xi_T$ and $S_T$ against $\chi$ appears robustly as the signatures of the criticality. The exponents $\kappa$ and the coefficients $\alpha$ change with the size. By substitution, the scaling relation between $S_T$ and $\xi_T$ [Fig. \ref{HCScaling} (c)] satisfies
\begin{eqnarray}\label{eq-scaling1}
S_T &=& c_T \ln \xi_T + \delta.
\end{eqnarray}
The factor $c_T = \alpha /\kappa$ is analog to the central charge, with $c_T\approx0.8$ for the sizes of $4 \times 4$ torus and $5 \times 5$ torus, and $c_T\approx1.1$ for $6 \times 6$ torus. The dependence of $c_T$ on different sizes might be caused by errors. Though the size-dependence of $c_T$ brings no harm to our main point which is using the scaling behaviors as signatures of criticality, we do believe it would be interesting and important to investigate whether $c_T$ relies on the size, and we show some preliminary results in the Appendix. This task requires more computational power and possibly some recently developed DMRG techniques, e.g., symmetries, to improve the efficiency.

\begin{figure}
	\begin{minipage}{\linewidth}
		\includegraphics[width=\textwidth]{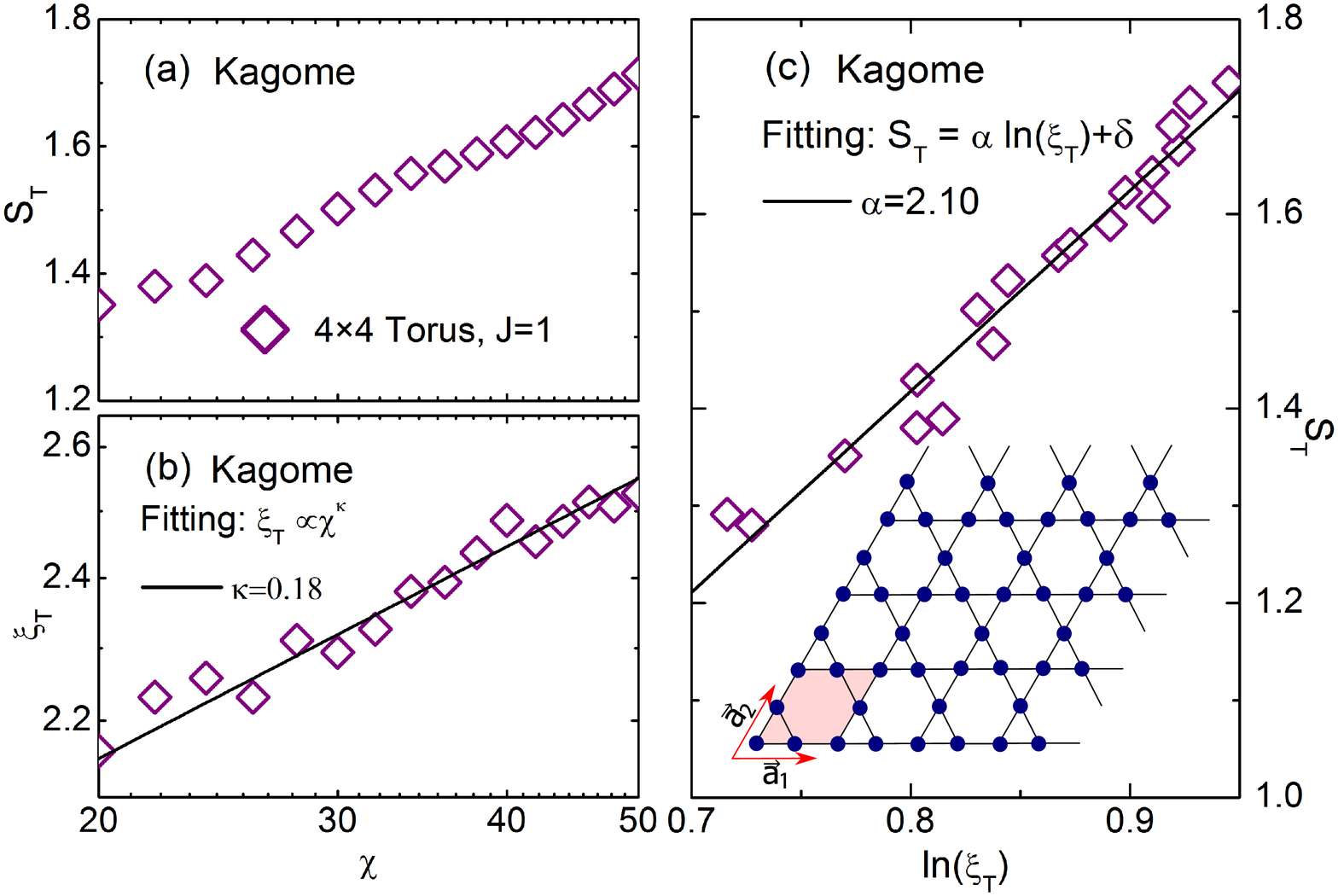}
	\end{minipage}
	\caption{\label{KLScaling}
		(Color online) The ground state scaling of (a) temporal entanglement entropy $S_T$ in the semi-log plot and (b) correlation length $\xi_T$ in the log-log plot of the spin-1/2 HAFM on kagome lattice. The $S_T$-$ln(\xi_T)$ scaling is in (c). The Hamiltonian reads $\hat{H}=J\sum_{<ij>}\hat{S}_i\hat{S}_j$, where $J$ is the isotropic coupling constant of the nearest neighbour spins $\hat{S_i}$ and $\hat{S_j}$. We fix $J=1$. The system on $4\times4$ torus is shown in the inset of (c).}
\end{figure}

We then apply the tMPS scaling method to study the ground state of the spin-1/2 HAFM on kagom\'e lattice. As shown in Fig. \ref{KLScaling}, the logarithmic relation between $S_T$ and $\chi$ and the algebraic relation between $\xi_T$ and $\chi$ are clearly observed. Our results are consistent with the experimental researches and some theoretical works where a gapless QSL is claimed \cite{INBSN08KagomeExp,HMSNB+07kagomeexp,OMBDTV+08KagomeExp,WLSRM+10kagomeexp,HHCNR+12fractionalized,HCBFZH+14KagomeGplessExp,RMAF07KagomeGapless,RHLW07kagomeQSL,HRLW08aQSL,SL09LowEnegExcKagome,IBSP13kagomeQSL,IPB14kagomeQSL,LXCLXH+17kagomeQSL,CRLPHS17arxiv,HZOP17kagomeDMRG}, and are against the results that predict a gapped $\mathds{Z}_2$ spin liquid \cite{IFHL11KagomeGappedExp,WEBLSL+98KagomeGapped,JWS08DMRGKagome,HT10KagomeGapped,DMS12DMRGKagome,YHW12DMRGkagome,JWB12TopoEntKagome,NSH12kagomemag,CDHLR13KagomePlateau}.

The advantage of our scheme is that the tMPS can faithfully give the criticality of 2D quantum systems even when the size is moderately small. This means that the dynamical correlations rely much less on the size than the spatial correlations. Our work also implies the existence of the 1D CFT in imaginary time as an alternative way to characterize the classes of 2D criticalities, being different from the Landau-Ginzburg paradigm with the critical exponents. Surely, more theoretical and numerical investigations are needed in the future.

% together show that the ground state of the HAFM on kagome lattice is critical. The reason comes from that the RG flow along the spacial path only maintain the 1D area law of entanglement entropy, and the correlation length convergence to a finite value because of suffering a strong finite size effect. Traditionally, one has to do the finite size scaling to investigate the 2D criticality. For gapped system, the truncation errors can be suppressed by improving the truncation dimension $\chi$ and enlarging the system size. While for gapless system, the entanglement spectrum becomes quit compact that prevents DMRG from expressing the real critical state by simply increasing the truncation dimension. From our result we are excited to find that the criticality is well defined by tMPS even though the system size is reasonably small.

\textit{Conclusion.---}
In this work, we propose an approach that can capture accurately the criticality of a many-body quantum system in two dimensions faithfully within small system sizes by introducing the tMPS, and utilizing the 2D NRG flow. We show the equivalence between the spatial MPS and tMPS in terms of correlation length and entanglement. We benchmark the scheme by calculating the entanglement entropy $S_T$ and correlation length $\xi_T$ of the spin-1/2 HAFM on honeycomb lattice; our results show that the isotropic and anisotropic cases obey quite distinct laws. For the former, $S_T$ has a logarithmic relation with the dimension cut-off $\chi$, while $\xi_T$ bears an algebraic decay law; for the latter, $S_T$ and $\xi_T$ saturate to a small value with increasing $\chi$, respectively. We apply the same  approach also to the spin-1/2 frustrated Heisenberg quantum antiferromagnet on kagom\'e lattice, and find that its ground state is gapless. The present study provides an efficient way in exploring the critical phenomena of quantum systems in two or even higher dimensions.

This work was supported in part by the MOST of China (Grants No. 2013CB933401), the NSFC (Grant No. 14474279), and the Strategic Priority Research Program of the Chinese Academy of Sciences (Grant No. XDB07010100). ML and SJR was supported by ERC AdG OSYRIS (ERC-2013-AdG Grant No. 339106), the Spanish MINECO grants FISICATEAMO (FIS2016-79508-P) and ``Severo Ochoa'' Programme (SEV-2015-0522), Catalan AGAUR SGR 874, Fundaci\'o Cellex, EU FETPRO QUIC (H2020-FETPROACT-2014 No. 641122)and EU grant EQuaM (FP7/2007-2013 Grant No. 323714), CERCA Programme / Generalitat de Catalunya.  SJR was supported by Fundaci\'o Catalunya - La Pedrera $\cdot$ Ignacio Cirac Program Chair.

\appendix

\setcounter{equation}{0}
\setcounter{figure}{0}
\setcounter{table}{0}
\makeatletter
\renewcommand{\theequation}{A\arabic{equation}}
\renewcommand{\thefigure}{A\arabic{figure}}
\renewcommand{\thetable}{A\arabic{table}}

\bigskip
\bigskip

\noindent \textbf{Appendix}

\section{Numerical renormalization flows and time matrix product states in the language of tensor network}
\begin{figure*}
    \begin{minipage}{\linewidth}
        \includegraphics[width=1\linewidth]{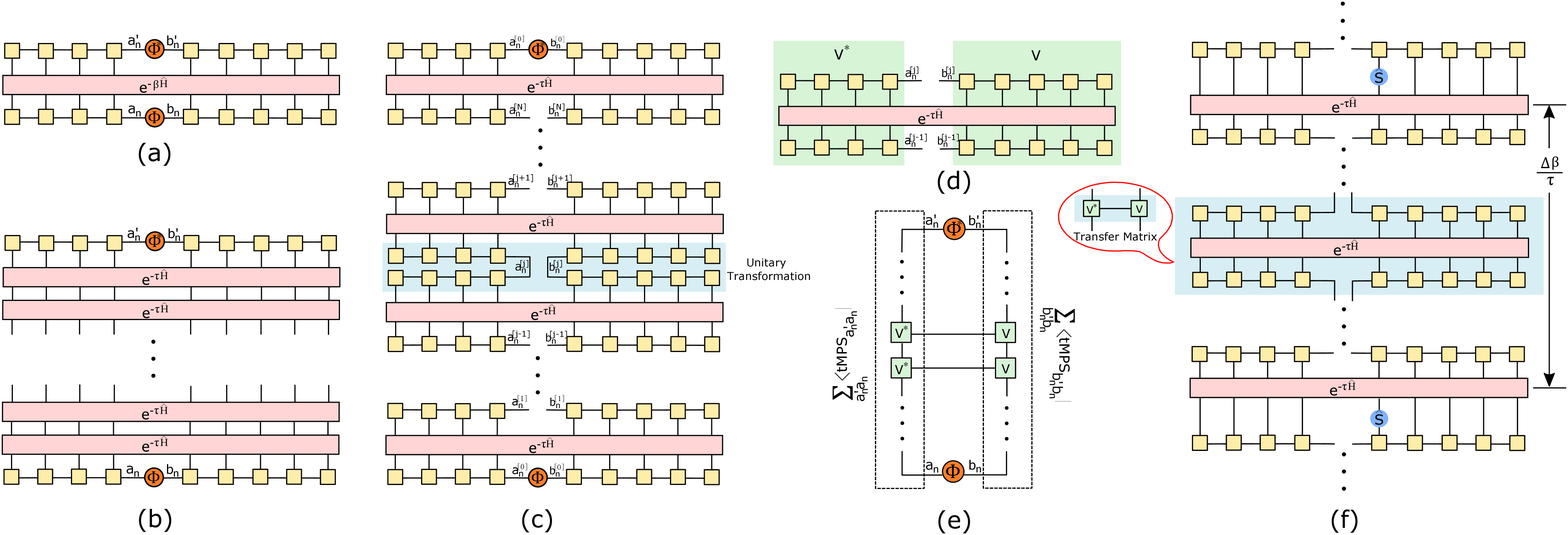}
    \end{minipage}
\caption{\label{tMPS}
(Color online) (a) The tensor network representation of $\langle \psi| e^{-\beta\hat{H}} |\psi \rangle$. The yellow squares are the local tensors of the MPS. $\Phi$ is the canonical center which carries the entanglement spectrum. The red rectangle is the matrix product operator (MPO) representation of $e^{-\beta\hat{H}}$ with $\beta$ the inverse temperature. (b) The Trotter-Suzuki decomposition of $e^{-\beta\hat{H}}$. (c) Unitary transformation (blue shaded region) of the tensor network in (b). (d) Decomposition of the effective Hamiltonian $H^{eff}_{a_n^{[j]}b_{n}^{[j]} a_n^{[j-1]}b_{n}^{[j-1]}}$. The green shaded regions are the local tensor $V^{\ast}_{c,a_n^{[j]} a_n^{[j-1]}}$ of $\sum_{a_n'a_n}\langle tMPS_{a_n'a_n}|$ and local tensor $V_{c,b_n^{[j]} b_n^{[j-1]}}$ of $\sum_{b_n'b_n}|tMPS_{b_n'b_n}\rangle$, respectively. (e) Reconstruction of the tensor network in (a) with tMPS. (f) The dynamic correlation function $\langle \hat{S}_{\beta}\hat{S}_{\beta+\Delta\beta} \rangle$ in the form tensor work. The blue circles are the spin operators $\hat{S}_{\beta}$ and $\hat{S}_{\beta+\Delta\beta}$. There are $\Delta\beta / \tau$ of the transfer matrices between the two operators.}
\end{figure*}

\begin{figure}
    \begin{minipage}{\linewidth}
        \includegraphics[width=\textwidth]{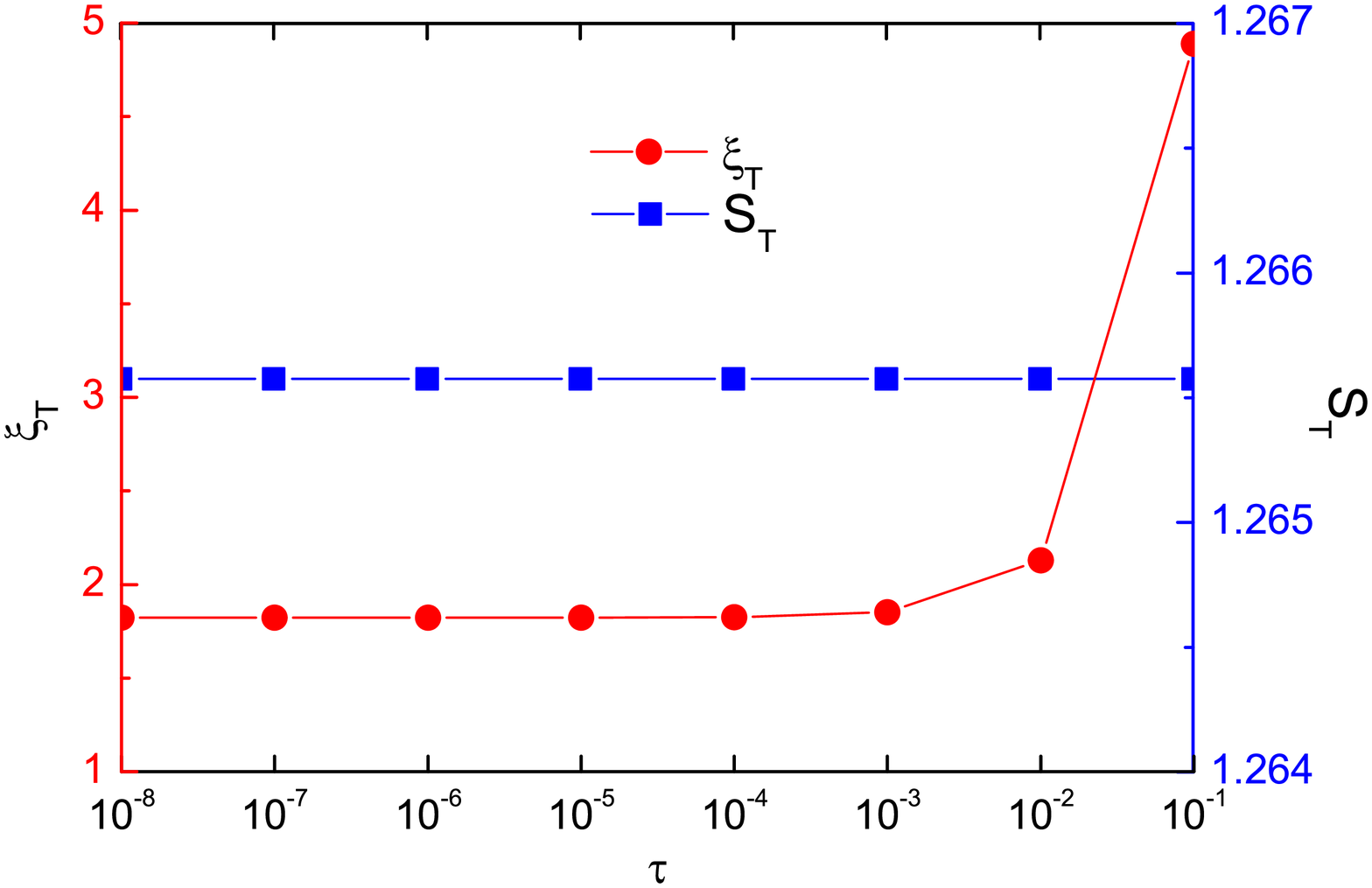}
    \end{minipage}
\caption{\label{Scalingtau}
(Color online) Trotter-Suzuki steps dependence of $S_T$ and $\xi_T$. Both of them converge quickly as decreasing the positive small number $\tau$, indicating that the Trotter error is neglected when $\tau$ is as small as $10^{-4}$. The results come from DMRG calculations implemented on the $4 \times 4$ torus of honeycomb lattice. The dimension cut-off is fixed as $\chi=30$.}
\end{figure}

Time matrix product state (tMPS) is a continuous matrix product state (MPS) along the imaginary time direction. In this section, we will explain in detail how to define the tMPS in a finite-size model using the language of tensor network (TN). This definition is applicable to the quantum systems in any dimensions. Most of the contents here are rephrasing and generalization of the work in Ref. [\onlinecite{TLR16tMPSarxiv}].

Let us take the 1D system of $N$ sites with the Hamiltonian $\hat{H}=\sum_{n}\hat{H}^{[n,n+r]}$ as an example. $\hat{H}^{[n,n+r]}$ gives the two-body interactions with $r$ the coupling distance. The ground state of the system can be represented in the form of MPS. The benefit of MPS is obvious: firstly, the parameter space needed for the ground state can be reduced from an exponential growth to polynomial as the system grows. Secondly, by introducing the canonicalization form, we can perform local operations without changing the irrelevant parts, thus saving considerable calculations. The variational ground state can be reached through various schemes, for example, to implement imaginary time evolution on a randomly initialized MPS to minimize the ground state energy as
\begin{equation}
\min \langle \psi| \hat{H} |\psi \rangle \rightleftharpoons \lim_{\beta\rightarrow\infty} \langle \psi| e^{-\beta \hat{H}} |\psi \rangle,
\label{apdxeq-min}
\end{equation}
where $\beta$ is the inverse of temperature. Practically, this procedure has to be equipped with Trotter-Suzuki decomposition because of the non-commuting two-body terms in the Hamiltonian $\hat{H}$. The right hand side of Eq.(\ref{apdxeq-min}) can be expressed utilizing the ground state MPS denoted in Eq. (\ref{eq-GSnew}) as
\begin{equation}
\langle \psi| e^{-\beta \hat{H}} |\psi \rangle = \sum_{a_n'b_n'a_nb_n} \langle \check{\psi}^L_{a_n}|  \Phi_{a_n'b_n'} \langle \check{\psi}^R_{b_n}| e^{-N\tau \hat{H}} |\check{\psi}^L_{a_n}\rangle  \Phi_{a_nb_n} |\check{\psi}^R_{b_n}\rangle,
\label{apdxeq-TSD}
\end{equation}
where $\tau$ is a small number, and $N$ needs to be extremely large so that $\beta=N\tau$ approaches to infinity.

%Here, we dub $e^{-\tau\hat{H}}$ as a Trotter slice.
The graphical representation of Eq.(\ref{apdxeq-TSD}) is shown in Figs. \ref{tMPS}(a) and (b). The ground state MPS is canonicalized for the canonical center $\Phi$ that carries the bipartite entanglement spectrum of the ground state. The dimensions of $a_n$ and $b_n$ are limited by the dimension cut-off $\chi$. The orthogonal conditions of the left and right NRG flows reads
\begin{eqnarray}
\langle \check{\psi}^L_{a_n'}|\check{\psi}^L_{a_n}\rangle &=& I_{a_n'a_n},\\
\langle \check{\psi}^R_{b_n'}|\check{\psi}^R_{b_n}\rangle &=& I_{b_n'b_n}.
\label{apdxeq-ortg}
\end{eqnarray}

By inserting (near) unitary transformation $\sum_{a_nb_n}|\check{\psi}^L_{a_n}\rangle\langle\check{\psi}^L_{a_n}| \otimes |\check{\psi}^R_{b_n}\rangle \langle\check{\psi}^R_{b_n}|$ between the Trotter slices as shown in Fig. \ref{tMPS}(c), we will not change the ground state spectrum. On the other hand, we find a series of effective Hamiltonian $\{H^{eff}_{a_n^{[j]}b_{n}^{[j]} a_n^{[j-1]}b_{n}^{[j-1]}}|j\in[1,N]\}$ according to the definition in Eq. (\ref{eq-Heffnew}). Note that all the physical dimensions are fused into either the left or the right NRG flow. If the decomposition in Eq. (\ref{eq-VLR}) exists (e.g., by performing the singular value decomposition), we can define the tMPS's $\sum_{b_n'b_n}|tMPS_{b_n'b_n}\rangle$ and $\sum_{a_n'a_n}\langle tMPS_{a_n'a_n}|$, respectively, constructed by the local tensors $V$ and its conjugate along the imaginary time direction (Figs. \ref{tMPS}(d) and (e)). Thus the effective Hamiltonian performs like the transfer matrix of $\sum_{a_n'b_n'a_nb_n}\langle tMPS_{a_n'a_n}|tMPS_{b_n'b_n}\rangle$. Finally, we can view the minimization problem from another angle, which is $\min \langle \psi| \hat{H} |\psi \rangle \rightleftharpoons \min \sum_{a_n'b_n'a_nb_n}\langle tMPS_{a_n'a_n}|\Phi^{\ast}_{a_n'b_n'}\Phi_{a_nb_n}|tMPS_{b_n'b_n} \rangle$. It is only true when $\Phi$ is the dominant eigenvector of the transfer matrix of the inner product of tMPS (Fig. \ref{tMPS} (e)).

We can see from above that the relationship between the imaginary time and space is inseparable. As explained in the main text, the bipartite entanglement spectrum of spatial MPS gives the temporal entanglement entropy $S_T$ of tMPS, and the inverse of the logarithmic gap of the transfer matrix is correlation length $\xi_T$.

The last but not the least, in Trotter-Suzuki scheme, one has to eliminate the Trotter error with infinitely small $\tau$. However, it is unrealistic to do so. Fortunately, both $S_T$ and $\xi_T$ converge in pace with decreasing Trotter-Suzuki steps (shown in Fig. \ref{Scalingtau}). This result tells us that we can choose a relatively small $\tau$, for example $\tau=1\times10^{-6}$ in our calculations.

\section{Density matrix renormalization group algorithm on two-dimensional lattices}
\begin{figure}
    \begin{minipage}{\linewidth}
        \includegraphics[width=0.8\textwidth]{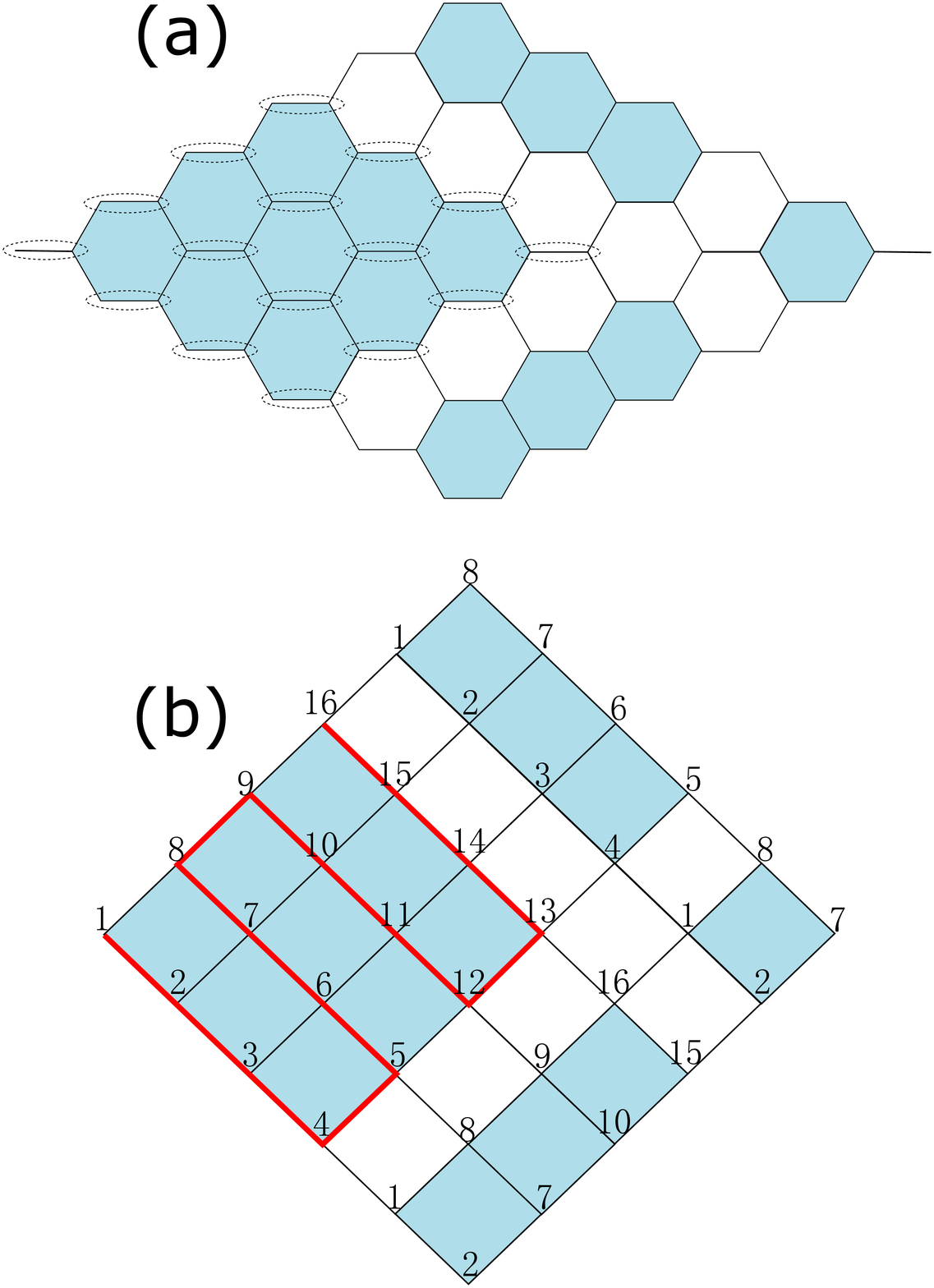}
    \end{minipage}
\caption{\label{Path}
(Color online) (a) $4\times4$ torus of honeycomb lattice. The sites in blue shaded parts break the translation invariance. Periodic boundary condition is represented by boundary interactions between different shaded parts. (b) The equivalent square lattice given by taking the two spins in the dashed ellipses of (a) as one site in the MPS. The number sequence marks each site in the S-shaped path.}
\end{figure}

\begin{figure}
    \begin{minipage}{\linewidth}
        \includegraphics[width=\textwidth]{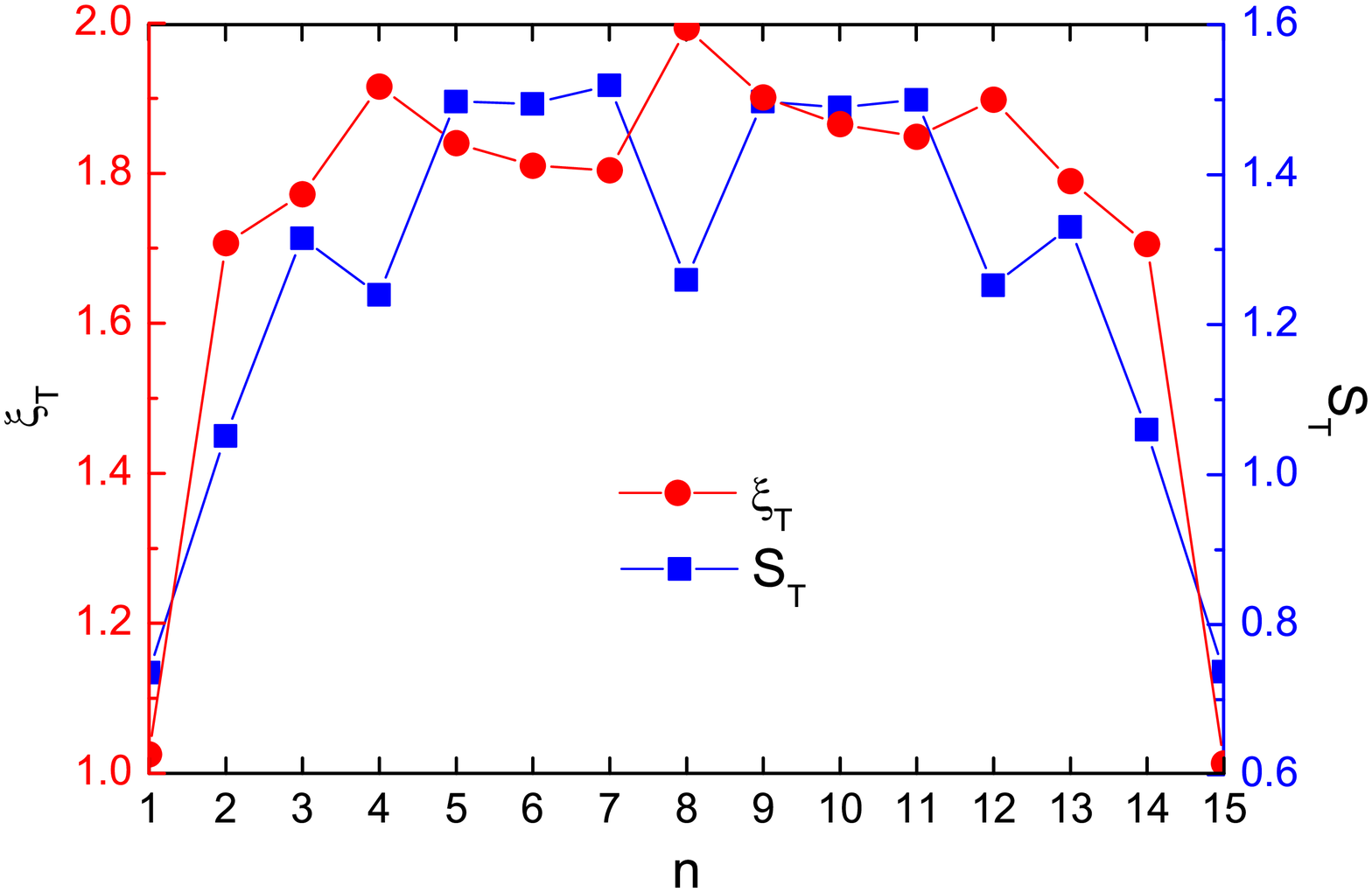}
    \end{minipage}
\caption{\label{ScalingSite}
(Color online) The position dependence of $S_T$ and $\xi_T$. The horizontal axis denotes the position of the canonical center of MPS. The dimension cut-off for DMRG calculations is fixed as $\chi=30$.}
\end{figure}

DMRG is a powerful tool of solving the ground state problem in 1D quantum strongly correlated systems. However in 2D, there exists systematic difficulties for DMRG in capturing the valid sub-space, i.e., to define 1D NRG flows for 2D states. The most common way is to choose a particular path on a 2D lattice. Thus the 2D system is transformed into a 1D model. It is inevitable that some nearest neighbor interactions become long-range ones in this procedure. As a consequence, one can only use DMRG in a lattice of finite size. For this reason, the scaling in the spatial direction is extremely challenging, including the scaling of the ground-state correlations and entanglement.

% Nevertheless, a compromising approach of considering an infinite large lattice is to put calculations on a small lattice equipped with periodic boundary conditions. Even though there is no need to require the shift invariance in the small lattice, the whole system is block translation invariant from the global perspective.

In the present work, we use DMRG to investigate the Heisenberg antiferrromagnet (HAFM) on honeycomb lattice and kagom\'e lattice. We take the block of honeycomb lattice shown in Fig. \ref{Path}(a) as an example. Spins are located on the sites; the interactions are between nearest neighbour spins, depicted by black lines. The blue shaded parts belong to the finite size honeycomb lattice. The black lines between different shaded parts denote the boundary interactions. In terms of the unit cell depicted in the inset of Fig. \ref{HCScaling}(c) in the main text, one can form a $4 \times 4$ torus. For convenience, we take the two spins in each dashed ellipse as one site in the MPS. Then, the honeycomb lattice virtually becomes a square lattice. We number the sites in the square lattice along a S-shaped path (shown in Fig. \ref{Path}(b)). Thus, there are $16$ inequivalent local tensors in the MPS. It is obvious that some physical nearest neighbor sites, for example, $1$ and $8$ are separated apart in the MPS construction.

As the sites on torus break the translation invariance, we should calculate the entanglement entropy $S_T$ and the correlation length $\xi_T$ at each virtual index of the ground state to see if all the sites are equivalent. From the calculated results in Fig. \ref{ScalingSite}, we identify certain fluctuations of those two quantities, which decrease when reaching the middle of the chain. In order not to lose generality, we use the average per site of $S_T$ and $\xi_T$ to do the scaling.

\section{Some results about the size-dependence of $\alpha$, $\kappa$ and $c_T$}
\begin{table}

\caption{\label{tab-sizedepen} Values of $\alpha$, $\kappa$ and $c_T$ obtained from fittings for the entanglement entropy $S_T$ and the correlation length $\xi_T$ on honeycomb lattice. The system sizes are chosen as $L \times L$ torus. The fitting functions refer to Eqs. (\ref{eq-scaling}) and (\ref{eq-scaling1}). The dimension cut-off $\chi$ ranges from $10$ to $70$ in our calculations.  }
\begin{tabular*}{8cm}{@{\extracolsep{\fill}}cccc}
  \hline\hline
  L & $\kappa$ & $\alpha$ & $c_T$ \\ \hline
  4 & 0.6(6) & 0.5(3) & 0.8(0)\\ \hline
  5 & 0.5(3) & 0.4(3) & 0.8(0)\\ \hline
  6 & 0.3(8) & 0.4(2) & 1.0(8)\\ \hline
  7 & 0.2(4) & 0.3(1) & 1.2(8)\\ \hline
  8 & 0.1(6) & 0.4(2) & 2.5(5)\\
  \hline\hline
\end{tabular*}

\end{table}

The size-dependence of the factors $\alpha$, $\kappa$ and $c_T$ are obtained by the fitting functions in Eqs. (\ref{eq-scaling}) and (\ref{eq-scaling1}). Here we present some preliminary results computed from the spin-1/2 HAFM on honeycomb lattice of different system sizes, as collected in Table. \ref{tab-sizedepen}. According to the present results, it seems that $\kappa$ decreases linearly as enlarging $L$ (the system size is $L \times L$ torus), while $\alpha$ converges to a constant, and $c_T$ goes up with increasing the size. Here we would like to mention that within the dimension cut-off that we can reach by our computing capacity, it appears that our calculations on large system sizes are less stable than those on small sizes, implying that more numerical attempts are needed in the future. Note that even though the size dependences of the factors are still to be unveiled, the scaling behaviors given by Eqs. (\ref{eq-scaling}) and (\ref{eq-scaling1}) are robust for all sizes we tried.

\end{document}